\documentclass{aa}
\usepackage{graphicx}
\usepackage{amsmath}
\begin{document}

\hyphenation{}

\title{Search for solid HDO in low-mass protostars}

\author{
   B. Parise\inst{1}
   \and T. Simon\inst{2}
   \and E. Caux\inst{1}
   \and E. Dartois\inst{3}
   \and C. Ceccarelli\inst{4}
   \and J. Rayner\inst{2}
   \and A.~G.~G.~M.~Tielens\inst{5}
}
\institute{
CESR CNRS-UPS, BP 4346, 31028 Toulouse Cedex 04, France
\and
Institute for Astronomy, 2680 Woodlawn Drive, Honolulu, HI 96822, USA
\and
IAS-CNRS, B\^at. 121, Universite Paris Sud, 91405 Orsay Cedex, France 
\and
Laboratoire d'Astrophysique, Observatoire de Grenoble, BP 53, 38041 Grenoble 
Cedex 09, France
\and
SRON, P.O. Box 800, NL-9700 AV Groningen, the Netherlands}
\offprints{Berengere.Parise@cesr.fr}

\date{Received {\today} /Accepted }
\titlerunning{Search for solid HDO in low-mass protostars}
\authorrunning{Parise et al.}

\abstract{We present ground-based 2.1 to 4.2 $\mu$m observations of four 
low-mass protostars. We searched for the 4.1~$\mu$m OD stretch band, characteristic 
of solid HDO in grain mantles. We did not detect solid HDO in any of the four 
sources, but we derive 3$\sigma$ upper limits from  
0.5\% to 2\% for the HDO/H$_2$O ratio depending on the source. 
These ratios provide strong constraints to solid-state deuteration models when compared to 
deuterium fractionation values observed in the gas phase. 
We discuss various scenarios that could lead to such a low water deuteration 
compared to the high formaldehyde and methanol deuteration observed in the gas-phase.
 }

\maketitle

\section{Introduction}

Large deuteration fractionations --- ratio of deuterated over normal 
isotope --- are a common characteristic of dark, 
dense molecular clouds, with abundances of deuterated species some 0.01-0.1 of
the normal isotopomer (e.g. \cite{Turner01} 2001). This deuteration is thought to reflect 
the small zero point vibrational energy differences between a deuterated 
species and its fully hydrogenated
counterpart, which drives the chemistry towards deuterated species
at low temperatures (10 K; \cite{Watson73} 1973).
In view of their elevated temperatures ($\ge$ 100 K), the warm gas around protostars 
shows unexpected high abundances of deuterated 
species. Fractionations of 10$^{-3}$ to more than 0.1 have previously been observed. 
Low-mass protostars seem to show even greater fractionations than 
high-mass protostars. Indeed, observations of the low-mass protostar 
IRAS 16293$-$2422 showed that D$_2$CO~/~H$_2$CO~$\sim$~10~\% 
(\cite{Ceccarelli98} 1998), a fractionation about 25 times larger than 
in the Hot Core in Orion (\cite{Turner90} 1990). 
Similarly large amounts of doubly-deuterated formaldehyde 
have subsequently been observed towards a sample of low-mass protostars 
(\cite{Loinard02} 2002). 
Singly- and doubly-deuterated methanol were detected towards IRAS16293$-$2422
(\cite{Parise02} 2002), showing fractionation ratios of 0.9 (CH$_2$DOH) and 
0.2 (CHD$_2$OH). Deuterated methanol was also observed towards a sample of 
other low-mass protostars (\cite{Parise03} in prep), confirming the huge 
deuteration in these objects, whereas the fractionation ratio 
CH$_2$DOH/CH$_3$OH is only 0.04 in the Orion high-mass star-forming region 
(\cite{Jacq93} 1993).

Because the high temperatures of the gas around protostars prevent 
significant deuterium fractionation, the observed high deuterations
are thought to reflect a previous cold phase. 
Molecules that formed during the dark cloud phase --- either in the
gas phase or on the grain surfaces --- are believed to be stored in an 
ice mantle on dust grains, which evaporates once the YSO heats its 
environment above the ice sublimation temperature (\cite{Ceccarelli01} 2001). 
 Presumably the current gas phase has not had time to return to
equilibrium, which takes some 3.10$^4$ years (\cite{Charnley92} 1992, 
\cite{Caselli93} 1993, \cite{Charnley97} 1997). 

A previous detection of solid-state HDO in the high-mass luminous W33A and 
NGC7538 IRS9 protostars has been claimed by \cite{Teixeira99} (1999). However, 
recent analysis of the same data seems to challenge this detection, giving 
an upper limit of 10$^{-2}$ for solid HDO/H$_2$O in NGC7538 IRS9
(\cite{Dartois03} 2003). Subsequent VLT observations of W33A also  
give an upper limit for the deuterium fractionation of $\sim$  10$^{-2}$ 
(\cite{Dartois03} 2003), a value still consistent with the typical deuteration 
observed for the gas 
phase of hot cores around 
high mass protostars. The search for HDO in high-mass protostars was thus 
thought to be inconclusive largely because the expected fractionation was 
very small. Because low-mass YSOs show gas-phase deuterium fractionations almost 100 
times higher in their hot cores, the solid HDO feature is expected to be much 
stronger, and motivated this study. Presently published grain chemistry models 
predict an HDO fractionation of about 40\% (i.e. almost all the deuterium 
would be locked in HDO) if deuterated water forms on the grain 
surfaces at the same time as formaldehyde and methanol (\cite{Caselli02} 2002, 
\cite{Stantcheva03} 2003).

In this paper, we present a search for the HDO stretch-band at 4.1 $\mu$m
towards four low-mass protostars. In section 2 we describe the observations,  
and derive the H$_2$O ice column density and upper limits for HDO column 
densities in section 3.
We discuss the derived fractionation ratios, comparing 
to available gas-phase observations (for SVS12 and SVS13) in section 4, and conclude in section 5.

\section{Observations}

\subsection{Source selection}

\begin{table*}
\begin{tabular}{cccccccccc}
\hline
\hline
Source  & $\alpha$ (2000.0) & $\delta$ (2000.0) 
 & m$_L$ & L-band  &  J-K & K & Slit / Resolution & On-source  & Standard \\
    &  & & & flux (Jy) && && int time & \\
\hline
NGC1333 / SVS12  & 03$^h$ 29$'$ 01.4$''$ & 31$^{\circ}$ 20$'$ 21$''$ &  && 4.24 & 10.61 & 0.3$''$ / 2500 &  20 min &  HD20995\\
NGC1333 / SVS13  & 03$^h$ 29$'$ 03.7$''$ & 31$^{\circ}$ 16$'$ 03$''$ & 5.4 &1.99& 3.74 & 8.25 & 0.3$''$ / 2500 &  15 min &  HD20995\\
L1489 IRS   & 04$^h$ 04$'$ 42.9$''$ & 26$^{\circ}$ 18$'$ 56$''$ & 6.8 &0.55& & 9.3& 0.3$''$ / 2500 &  20 min &  HD23258 \\ 
TMR1  & 04$^h$ 39$'$ 13.9$''$ & 25$^{\circ}$ 53$'$ 21$''$  & 8.9 &0.08& 5.55 & 10.54 & 0.5$''$ / 1500 &  53 min &  HD31592\\
\hline
\end{tabular}
\caption{Observed sources, coordinates, NIR magnitude, J-K colour index (indicative of extinction, cf text), K magnitude and observation parameters.}
\label{sources}
\end{table*}

As noted in the Introduction, the previous searches for solid HDO have been 
carried out towards high-mass protostars, with very modest success.
In this study we focused on low-mass protostars, where gas-phase deuteration 
has been observed to be much higher than in high-mass protostars.

The sources of the present study
were selected to be bright enough at NIR wavelengths, 
so that high enough S/N can be obtained to detect weak absorption features against the continuum, and 
to have large J-K colour index, which would indicate a high extinction.

The first criterion, by definition, excludes the most embedded low-mass 
protostars, the Class 0 sources, because the NIR continuum is too faint, leaving
the more evolved Class I sources. Even though not totally conclusive, 
the study of \cite{Loinard02} (2002) on a sample of four Class I sources 
seems to indicate that also in those sources the molecular deuteration, and 
specifically the D$_2$CO/H$_2$CO ratio, remains relatively large and 
comparable to that found in Class 0 sources. 
Encouraged by this, we selected the four sources in Table \ref{sources}. 
Two of our sources were observed by \cite{Loinard02} (2002).
In the following we give a brief description of each selected source.

{\bf NGC1333 SVS12}, also called IRAS6, 
is probably a 
Class I source of $\sim$ 28 L$_\odot$ in the IRAS bands 
(\cite{Jennings87} 1987) assuming a distance of 350 pc. New distance 
estimates 
of the NGC1333 complex tend to put it closer at about 220 pc 
(\cite{Cernis90} 1990), which would bring the SVS12 luminosity to 
$\sim$ 10 L$_\odot$.
The infrared source is close to the Herbig Haro object HH12, 
although SVS12 probably is not the 
exciting source of the HH12 flow. Millimetre and submillimetre maps resolved 
the source in possibly three components (\cite{Sandell01} 2001). A detailed 
CO 2-1 map of the region did not detect any outflow emanating from this 
source (\cite{Knee00} 2000) so it may be a relatively evolved source indeed,
and possibly a background Class II source. 
In this case, we may be observing the 
deuteration in the cloud ices rather than in the circumstellar material.
\cite{Loinard02} (2002) detected abundant doubly deuterated formaldehyde,
D$_2$CO/H$_2$CO $\sim 5$\%.


{\bf NGC1333 SVS13}, also called IRAS03259+3105 (\cite{Jennings87} 1987), is the 
best studied source of the region, thanks to its well collimated flow of 
HH objects, the HH7-11 complex. It is a variable object and one of the 
brightest NIR sources of the 
region, having an average bolometric luminosity of 115 L$_\odot$ (\cite{Molinari93} 1993;
45 L$_\odot$ if the distance is 220 pc), 
and it is very probably a multiple system (\cite{Lefloch98a} 1998a, b; 
\cite{Bachiller98} 1998; \cite{Looney00} 2000). 
Also in this source \cite{Loinard02} (2002) detected abundant doubly 
deuterated formaldehyde, D$_2$CO/H$_2$CO $\sim 4$\%.

{\bf L1489 IRS} (IRAS04016+2610) is a low-luminosity source (L$_{bol}$=3.7 
L$_\odot$),
which acquired recent attention because it is suspected to be one of the rare 
cases of a source in transition from Class I to Class II 
(\cite{Hogerheijde00} 2000; \cite{Hogerheijde01} 2001). It appears to be 
surrounded by a relatively massive and young disk (\cite{Boogert02a} 2002a), 
whereas the envelope seems to be largely swept out.

{\bf TMR1} (IRAS04361+2547) is a prototypical Class~I, low-luminosity source 
(L$_{bol}$=2.9 L$_\odot$:   \cite{Hogerheijde98} 1998).
Both L1489 IRS and TMR1 belong to the Taurus complex at a distance of 140 pc.

\begin{figure*}
\includegraphics{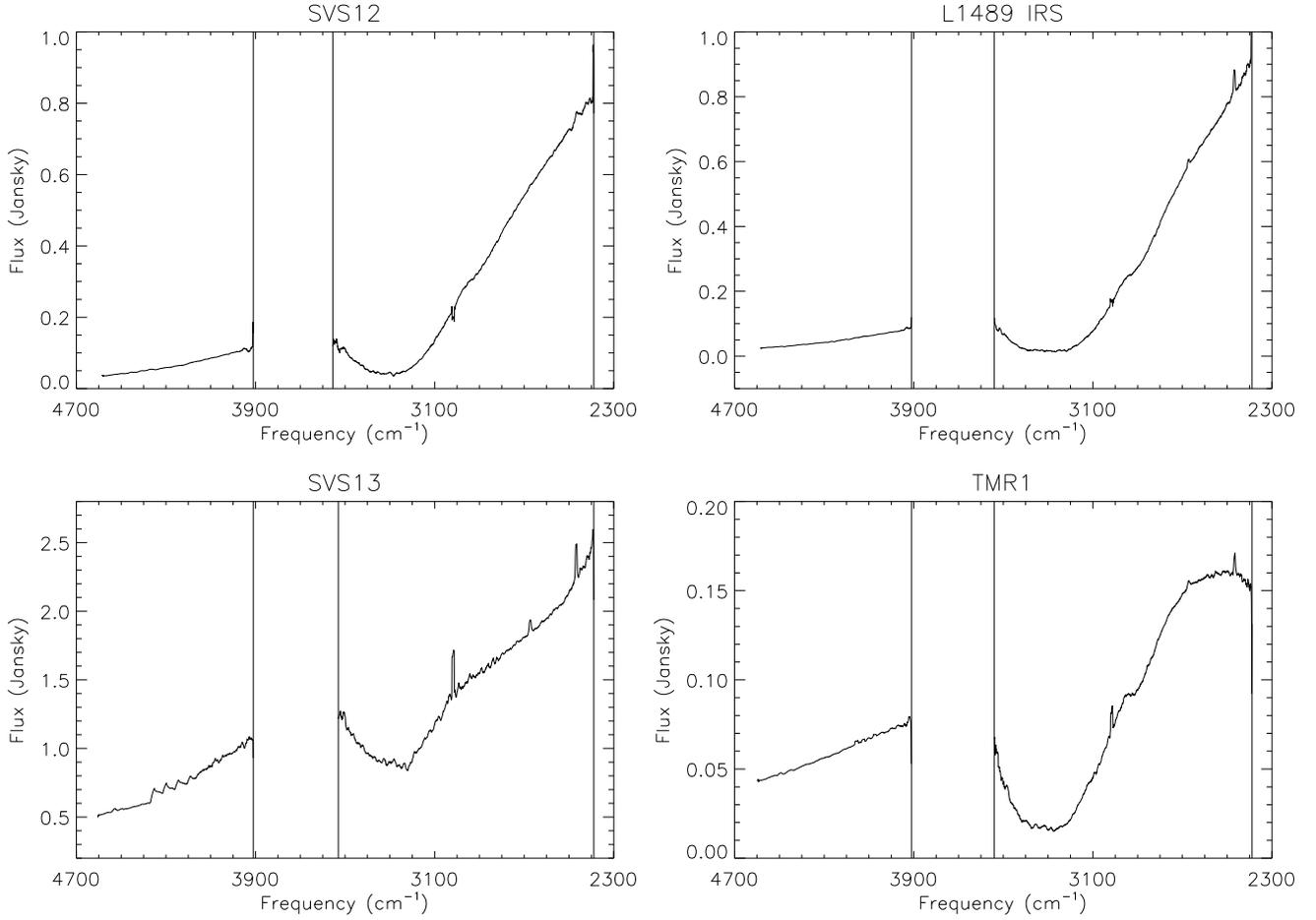}
\caption{Spectra of the four protostars, smoothed to 10 cm$^{-1}$ resolution. The vertical lines delineate the region of strong atmospheric absorption features. }
\label{obs}
\end{figure*}

\subsection{Observations}

Using the SpeX instrument (\cite{Rayner03} 2003) on the IRTF telescope on 
Mauna Kea (Hawaii), 
we observed the four low-mass class I protostars listed in Table \ref{sources}.
The observations were performed in December 2002 under conditions of good 
seeing (0.6$''$ at K, on average).
We used the 2.1-5.0 $\mu$m cross-dispersed mode, which acquires the full 
spectrum between 
2.1 and 5.0 $\mu$m simultaneously. We used the 0.3$''$ slit, which provides a spectral 
resolution of 2500, for the first three objects, and the 0.5$''$ slit (R=1500) 
for TMR1. Standard stars were chosen from the IRTF A0V-star
database to be as bright and as close to the studied object as possible, and are 
listed in Table \ref{sources}. The airmass difference between the object 
and the associated standard was always less than 0.1.

The observations were reduced using the Spextool software (\cite{Vacca03} 2003,
Cushing et al. in prep), which is available 
on the IRTF SpeX webpage (http://irtfweb.ifa.hawaii.edu/Facility/spex/).
The full spectra for the four protostars, smoothed to a resolution of $\sim$ 10 cm$^{-1}$, 
are presented in Fig. \ref{obs}.

The HDO absorption feature was not detected in any of the four sources. We 
derive upper limits for the HDO/H$_2$O ratio in next section.
The spectra reveal two hydrogen lines at 2468 cm$^{-1}$ (4.05$\mu$m) and 2673 cm$^{-1}$ (3.74$\mu$m), as well 
as CO emission lines (SVS13, \cite{Carr92} 1992). The deep absorption feature at 3300 cm$^{-1}$ (3.03$\mu$m)
is the OH stretch band. On its right wing, the so-called ``3.47 $\mu$m feature'' (2882 cm$^{-1}$) is visible.   

\section{Water features}

\subsection{The OH stretch band}

All four spectra show a prominent OH absorption feature at 3300~cm$^{-1}$ 
(3.03~$\mu$m). The abundance of solid hydrogenated water can be derived 
from this OH stretch band. We first define the continuum I$_{cont}$ around 
the OH feature.

We can then plot the optical depth $\tau$ defined by :

$$ I / I_{cont} = e^{-\tau}. $$

The column density of H$_2$O molecules can be derived from the optical 
depth with the following relation~:
$$ N_{H_2O} = \frac{\int \tau d\nu}{A_{H_2O}},$$ 
where the band strength $A_{H_2O} = 2\times10^{-16}$ cm/molecule 
(\cite{Dartois98} 1998) and  $\int \tau d\nu$ is expressed in cm$^{-1}$.

\subsubsection{SVS12, L1489 IRS and TMR1}

\begin{figure}
\includegraphics[width=9cm]{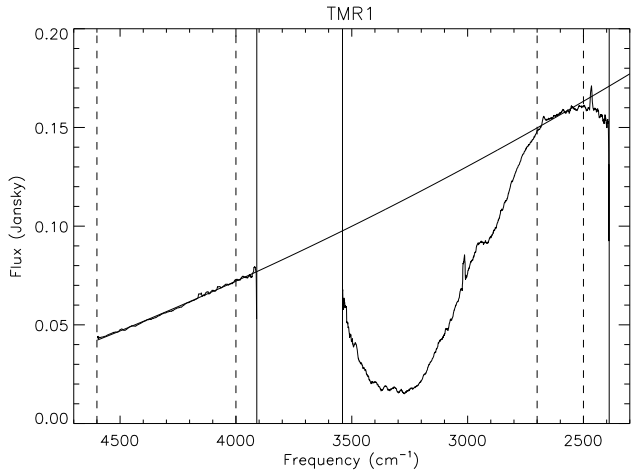}
\caption{Example of determination of the continuum for the study of the 3.0$\mu$m feature. The dotted lines show the frequency ranges used to fit the continuum}
\label{continuumh2o}
\end{figure}

For these three sources, we defined the local continuum around the OH feature
by fitting a second order polynomial to the frequency ranges [2400-2500, 2700]
and [4000, 4400-4600] cm$^{-1}$. These frequency ranges were chosen to avoid the 
region around the 2882~cm$^{-1}$ (3.47~$\mu$m) feature in
the continuum fit. 
An example of the derived continuum is shown in Fig.~\ref{continuumh2o}.

\begin{figure}
\includegraphics[width=9cm]{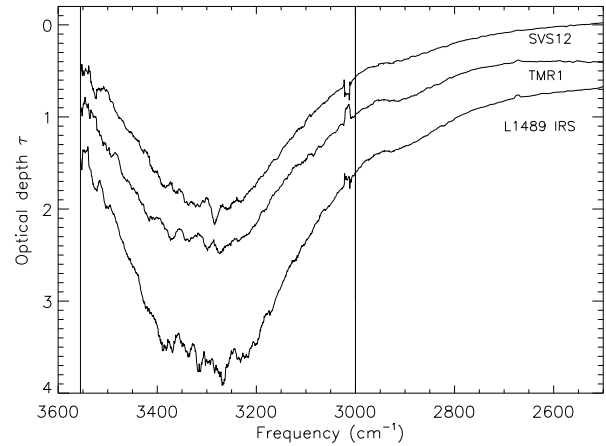}
\caption{OH stretch band for SVS12, L1489 IRS and TMR1. For clarity, the plots for TMR1 and L1489 IRS have been offset by 0.4 and 0.7 respectively. The vertical lines delineate the band
where the integrated optical depth was computed.}
\label{h2o}
\end{figure}

The water feature for the three sources shows no structure and is thus attributable
to amorphous ice. We integrated the contribution of the band between 3000 and 3555 cm$^{-1}$. 
This integrated optical depth corresponds to a lower limit on the H$_2$O column density.  
The OH bands for these three sources are plot in an optical depth scale on Fig. \ref{h2o}. 

The optical depth for the 3300~cm$^{-1}$ feature for each source and the 
derived H$_2$O column density are summarized in Table \ref{coldens}.
Our optical depth for L1489 IRS is consistent with the observation
at low resolution (R=150) by \cite{Sato90} (1990; $\tau = 2.9 \pm 0.2$), 
but we find a lower (2.0) optical depth for SVS12 ($\tau = 2.9 \pm 0.2$, 
\cite{Sato90} 1990). This inconsistency certainly arises from the continuum 
determination, as \cite{Sato90} only had observations between 2.4~$\mu$m 
(4167~cm$^{-1}$) and 3.8~$\mu$m (2632~cm$^{-1}$).

\subsubsection{SVS13}
\label{SVS13}
 
SVS13 appears to be a more unusual source than the other three. First of 
all, its OH stretch band has a non-gaussian shape, which suggests that the ice 
is in a crystalline state. Second, 
the spectrum is characterized by a strange inflexion in the frequency range 
[3900, 4700] as well as
[2400, 2600] cm$^{-1}$. The inflexion in the [2400, 2600] cm$^{-1}$ range 
cannot come from the standard star, as we used the same standard object 
for SVS12, which shows no inflexion. Moreover, 
ISO-SWS observations at low resolution of this source 
(ISO archive tdt 65201959) have the same inflexion. This strange behaviour 
of the spectrum makes it difficult to define the local continuum.

\begin{figure}
\includegraphics[width=9cm]{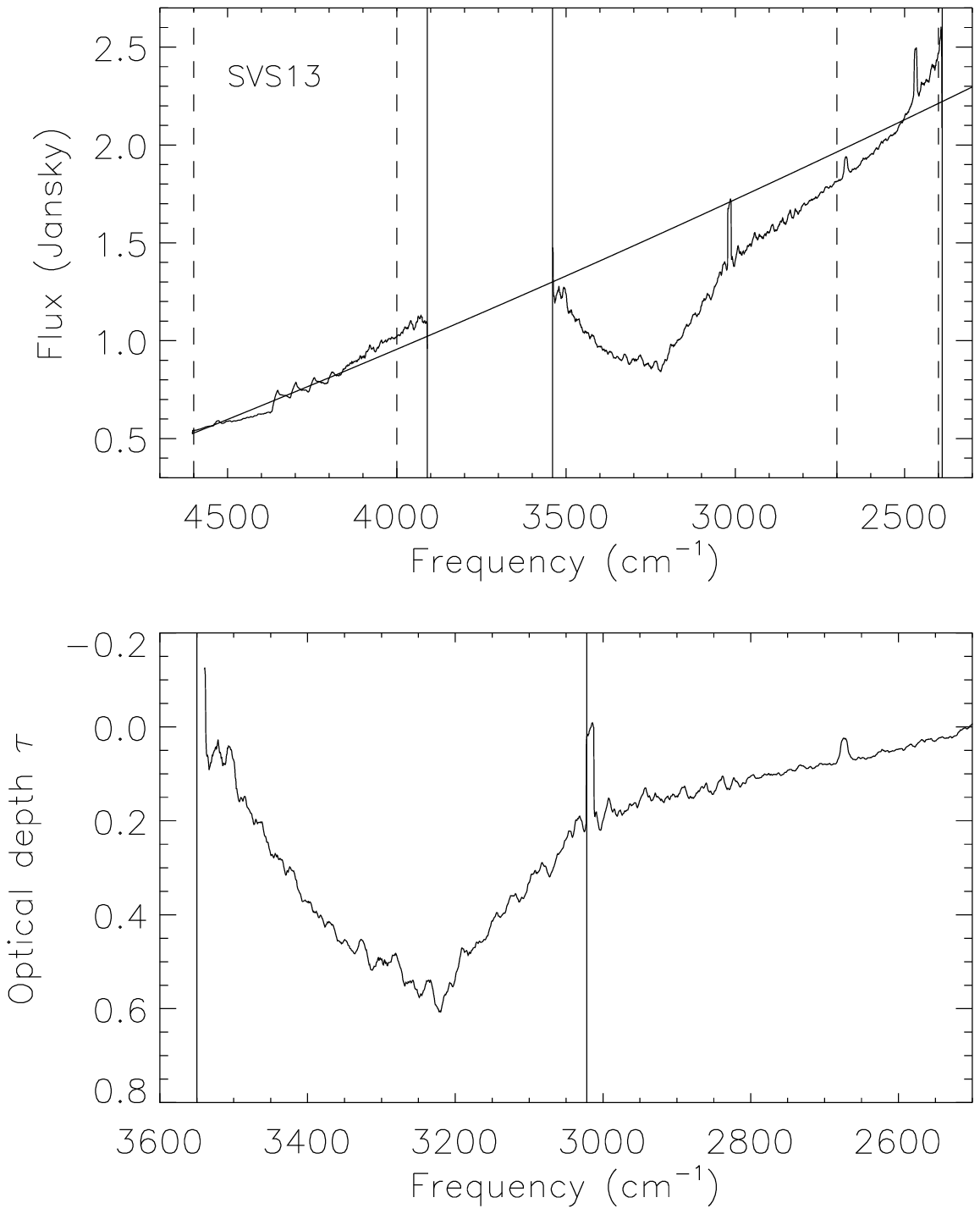}
\caption{Example of determination of the continuum for the study of the 
3.0$\mu$m feature. The dotted lines show the frequency ranges used to fit 
the continuum.}
\label{SVS13continuum}
\end{figure}

We defined several different continua, by fitting first order polynomials on different 
anchoring frequency ranges. An example is shown in Fig. \ref{SVS13continuum}. 
In all cases, the optical depth plots
have a broad absorption on the right side of the OH feature. This absorption 
could be due to scattering in the OH wing (\cite{Dartois02} 2002).
The integrated optical depth varies from 190 to 220 cm$^{-1}$. 
We chose the smallest value for a lower limit on the H$_2$O column density.

\begin{table*}
\begin{center}
\begin{tabular}{cccccccc}
\hline
\hline
       &       & $\int \tau_{OH} d\nu$ & N(H$_2$O) & &$\int \tau_{OD} d\nu$ & N(HDO) & HDO/H$_2$O \\
Source & $\tau_{OH}$&  (cm$^{-1}$)     & (10$^{18}$ cm$^{-2}$) &$\tau_{OD}$ & (cm$^{-1}$)     & (10$^{16}$ cm$^{-2}$) & \\
\hline
NGC1333 / SVS12 & 2.0 & $\ge$ 750 & $\ge$ 3.8 & $\le$ 0.005  & $\le$ 0.63 & $\le$ 1.8 & $\le$ 0.5 \%\\
NGC1333 / SVS13(a) & 0.55 & $\ge$ 190 & $\ge$ 1.0 & $\le$ 0.016  & $\le$ 2.0 & $\le$ 5.6 & $\le$ 5.9 \% \\
NGC1333 / SVS13(b) & & & &$\le$ 0.022  & $\le$ 0.58 & $\le$ 1.6 & $\le$ 1.7 \% \\
L1489 IRS  & 3.0 & $\ge$ 1160 & $\ge$ 5.8 & $\le$ 0.013  & $\le$ 1.6 & $\le$ 4.4 & $\le$ 0.8 \% \\
TMR1  & 2.0 & $\ge$ 770 & $\ge$ 3.9 & $\le$ 0.013  & $\le$ 1.6 & $\le$ 4.4 & $\le$ 1.1 \% \\
\hline
\end{tabular}
\caption{OH and OD stretch band characteristics. Upper limits for HDO parameters are 3-$\sigma$. (a) in the amorphous case, (b) in the crystalline case (see text).}
\label{coldens}
\end{center}
\end{table*}

\subsection{Upper limits on the HDO abundance}

\begin{figure*}
\includegraphics{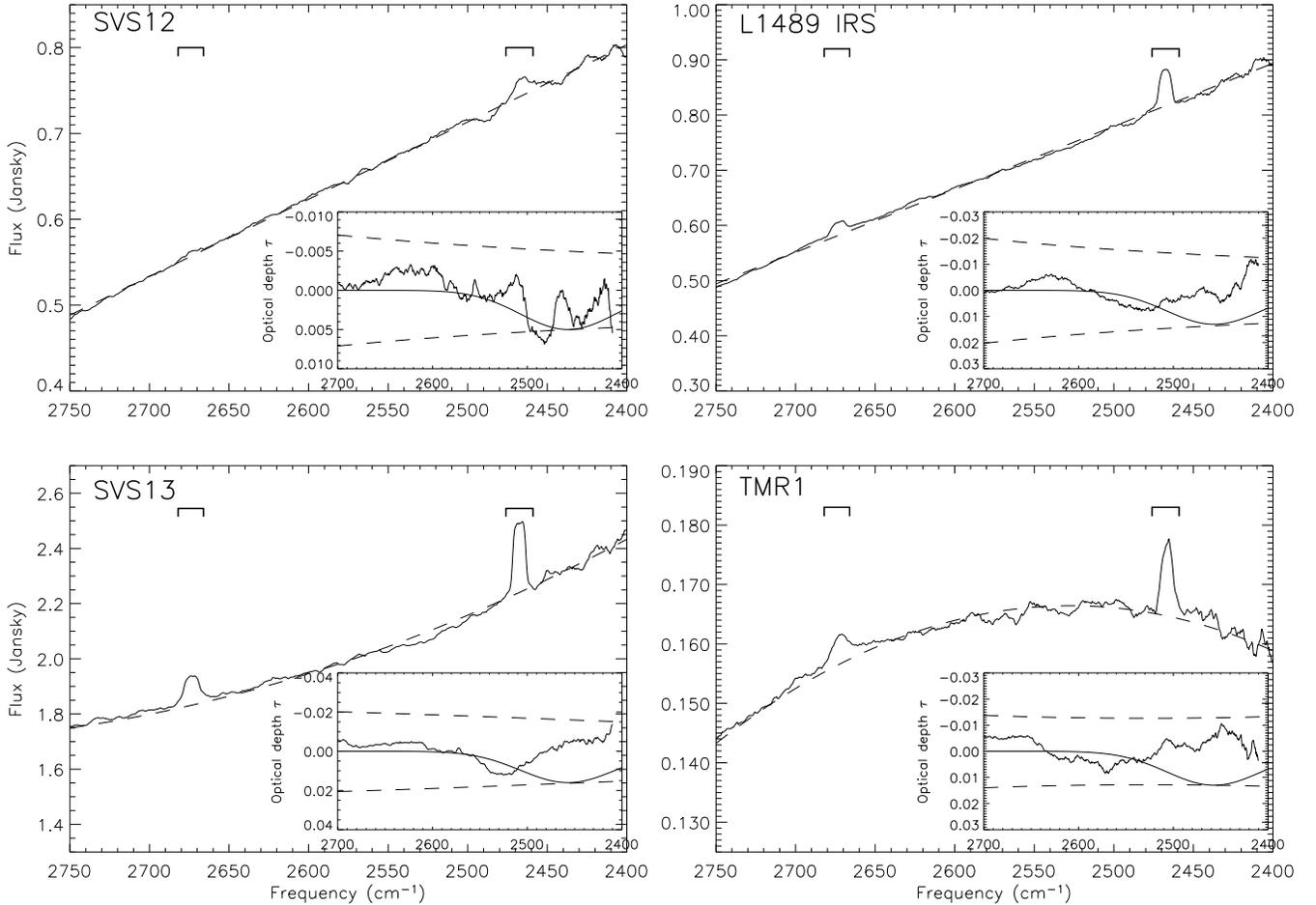}
\caption{OD stretch band for the four protostars. The horizontal bars show 
the frequency ranges ignored in the continuum fit. The continuum is drawn as 
a dashed line. In insets, the data are plotted on an optical depth scale versus
 frequency, after smoothing to 30 cm$^{-1}$ resolution. The dashed lines are 
the 3$\sigma$ limits on the flux, transposed on an optical depth scale. The 
gaussian shows the maximum amorphous HDO absorption.}
\label{HDOfit}
\end{figure*}

To determine upper limits on the HDO abundances, we first subtracted the two 
hydrogen lines at 2468 and 2673~cm$^{-1}$. To do so, we fitted a first or second 
order polynomial in the [2400, 2800] cm$^{-1}$ frequency range, ignoring a small frequency 
range around each of the two hydrogen lines (horizontal marks in Figure \ref{HDOfit}). 
Continuum fits are plotted for each source in Figure \ref{HDOfit}.
The hydrogen lines were subtracted from the data by replacing the data 
in the associated frequency ranges by the interpolated continuum. 
   
We then defined a local continuum 
around the expected 2457~cm$^{-1}$ (4.07$\mu$m) OD stretch band by fitting 
a first or second order polynomial in the [2410,2440] and [2500,2800] cm$^{-1}$ 
frequency range.

An upper limit on the HDO column density was derived taking into account 
that the expected feature is rather broad (120~cm$^{-1}$). 
The data were smoothed accordingly to a resolution of 30~cm$^{-1}$. The 
upper limit on the HDO column density 
was then derived by estimating the biggest gaussian feature fitting in a 
3$\sigma$-wide band around the polynomial fit.
 
The gaussian, centered on 2457 cm$^{-1}$, was chosen to have a fixed FWHM of 
120 cm$^{-1}$ (0.2 $\mu$m), as measured in the laboratory for amorphous HDO 
(\cite{Dartois03} 2003). The extreme gaussians for each source are plotted on 
an optical depth scale in Figure \ref{HDOfit} (see insets). The dashed lines represent 
the $\pm$ 3$\sigma$ limits associated with the continuum determination at 
30~cm$^{-1}$ resolution.


The upper limit on the HDO column density is then derived from the relation :
$$ N_{HDO} \le \frac{\int \tau d\nu}{A_{OD}} $$ where $\int \tau d\nu$
is the area of the extreme gaussian, and $A_{OD}$ was taken to be $3.6\times10^{-17}$
for a conservative upper limit estimate ($A_{OD} = (4.3\pm0.7)\times10^{-17}$cm/molecule, \cite{Dartois03} 2003). 

\begin{figure}
\includegraphics[width=9cm]{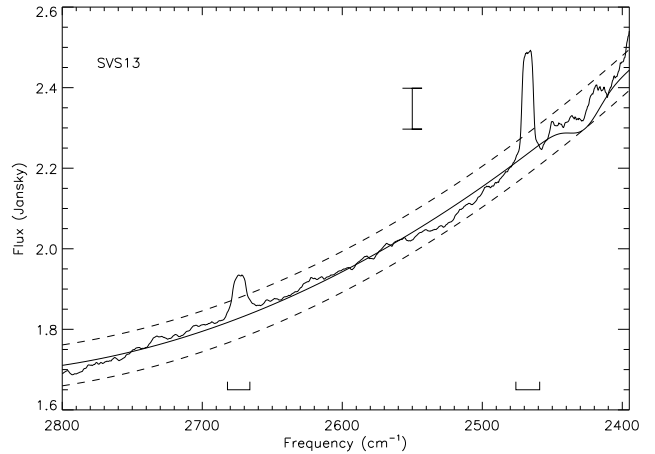}
\caption{OD stretch band for SVS13. The horizontal bars show the frequency 
ranges ignored in the continuum fit. The vertical bar is a representative 
$\pm$ 3$\sigma$ error bar. The maximum gaussian absorption, superimposed as 
a solid line, has been computed under the assumption of crystalline ice 
(see text).}
\label{SVS13_HDOcris}
\end{figure}

For SVS13, we estimated an upper limit on the HDO column density for 
the case of crystalline ice. The gaussian was then centered on 
2427~cm$^{-1}$ (4.12~$\mu$m), with a FWHM of 30~cm$^{-1}$ (0.05~$\mu$m) 
following laboratory experiments (\cite{Dartois03} 2003). In that case, the data 
were smoothed to 10~cm$^{-1}$ resolution, the results for which are plotted on 
Figure \ref{SVS13_HDOcris}.

The 3$\sigma$ upper limits for the HDO column densities and
associated HDO/H$_2$O ratios are listed in Table \ref{coldens}.

\section{Discussion}

 The observations presented here represent the first attempt to study the
solid-phase deuteration in low-mass protostars, namely sources 
with luminosities of a few tens of solar luminosities.
Previous studies focused on sources whose luminosity is larger than
about 200 L$_\odot$ (e.g. Dartois et al. 2003), and where the deuteration
of gas phase molecules is relatively low.
Specifically, the gas-phase D$_2$CO/H$_2$CO ratio is  $\leq 0.4$\% in high-mass 
protostars, whereas it is $\geq 4$\% in low-mass protostars
(\cite{Loinard02} 2002).
For this reason, the present observations represent the most
stringent constraint and a crucial test 
on the relation between the solid H$_2$O deuteration and the
large deuteration observed in the minor gas-phase species.

Loinard et al. (2002) reported the observation of D$_2$CO/H$_2$CO
in two (SVS12 and SVS13) out of the four sources studied here.
In both sources the measured D$_2$CO/H$_2$CO ratio is around 4\%,
a value similar to that found also in the other two Class~I sources
studied by \cite{Loinard02} (2002), and just a factor two 
or three smaller than the average value measured in Class~0 sources.
Hence the case of SVS12 and SVS13 is likely
representative of the formaldehyde deuteration in Class~I sources,
and more generally in embedded low-mass protostars, within a factor two 
or three.
Note that the HDCO/H$_2$CO ratio in embedded protostars
is even larger, $\sim 10$\%, as
observed in the Class 0 source IRAS16293-2422 (\cite{Loinard00} 2000, 
\cite{vanDishoeck95} 1995).
Not to mention the case of the deuterated methanol, which is about 
as abundant as the main isotope (\cite{Parise02} 2002).

The present observations clearly show that in low-mass protostars
the solid HDO/H$_2$O ratio is lower than about 2\% (3-$\sigma$), 
definitively lower than the observed fractionation 
of the gas-phase formaldehyde. This is a crucial step towards the understanding
of the deuteration from an observational point of view.

Before discussing the consequences of the comparison of these ice observations
with gas-phase observations, it is important to understand where these ices are located 
along the line of sight and whether the ices 
we are probing have experienced the same chemical history as the ices that evaporated
in the hot core and released the high-deuterated molecules such as formaldehyde and 
methanol into the gas-phase. In principle, a substantial part of the ice absorption 
may arise in the foreground quiescent gas not directly associated with the protostar 
itself (\cite{Boogert02b} 2002b). Observation of the pure solid CO band at 
4.67~$\mu$m may help elucidate how 
much thermal processing the ices have undergone and hence their proximity to the YSO. 
Indeed, pure CO absorption will only be present for ices less than 15~K and hence is 
expected to arise only in the dark foreground cloud. The CO band towards L1489 IRS 
has been observed and modelled by \cite{Boogert02a} (2002a). They find an optical depth of 
0.4 for the pure CO component (see their figure 5). The distribution of pure solid CO
in the Taurus molecular cloud has been well studied (\cite{Whittet89} 1989, 
\cite{Chiar95} 1995). The optical depth in the unperturbed dark cloud scales 
linearly with the total extinction ($\tau_{pure CO}=0.08 (A_v-6)$, \cite{Chiar95} 1995).
The pure CO optical depth towards L1489~IRS corresponds then to a visual extinction
of 11 for the foreground cloud and a visual extinction of 18 for the envelope associated
with the protostar. The $H_2O$ ice optical depth in the Taurus dark cloud shows a 
similar linear correlation with A$_v$ ($\tau_{H_2O}=0.08 (A_v-3)$, 
\cite{Chiar95} 1995). Hence, the H$_2$O optical depth associated with the foreground 
cloud is only 0.6 and ice grain in the envelop of L1489~IRS contribute 
$\tau_{H_2O,ice}=2.4$. Thus, our upper limit on the fractionation of H$_2$O ice 
in the immediate environment of this protostar is 1\%.

In the following we discuss the consequences of these two sets of
observations, gas-phase and solid-state deuteration.
We start discussing two possibilities: either formaldehyde (and methanol) forms at the
same time as water ices or water ices are formed in a previous phase
and therefore they are imprinted with the --different--
deuteration degree of this phase.

There is now a general consensus, supported by several observational
and theoretical works, that both formaldehyde and methanol
form on the grain surfaces, probably after hydrogenation of CO molecules that have stuck 
onto the grains (e.g. theoretical studies of \cite{Shalabiea94} 1994; \cite{Tielens97} 
1997; \cite{Charnley97} 1997, and laboratory experiment of \cite{Watanabe02} 2002), although 
one laboratory experiment seems to challenge this last hypothesis (\cite{Hiraoka02} 2002). 
If we assume now that water ices are formed at the same time by hydrogenation
of oxygen atoms sticking on the grain surfaces, the deuteration 
degree is set by the atomic D/H ratio of the accreting gas (e.g. \cite{Tielens83} 
1983; \cite{Caselli02} 2002), and cannot be substantially different in 
the three species (water, formaldehyde and methanol), unless selective deuteration 
is at work in formaldehyde and methanol with respect to water. Detailed chemical 
models have been developed for the deuterium fractionation on grain surfaces. These 
lead to enhanced fractionation of formaldehyde with respect to that in other ice 
species centering on H-abstraction reactions, driven by the zero-point energy 
difference (\cite{Tielens83} 1983). However, our insights in grain surface 
chemistry have evolved considerably since then and these models have not been 
fully revisited.

In the second hypothesis, water, formaldehyde and methanol would be 
formed during different phases characterized by different physical 
conditions which then may lead to different levels of fractionation 
(cf., \cite{Dartois03} 2003).  In particular, if the methanol and 
formaldehyde are formed during the later stages of accretion - when most 
of the gas has accreted already --, their fractionation could 
be high (\cite{Roberts03} 2003).  Such a model would require that the 
physical conditions in the accreting gas which favor deuteration also 
favor formaldehyde and methanol formation but not water formation. 
This hypothesis encounters some problems as the high abundance 
of formaldehyde and methanol in hot cores is generally thought to 
reflect the evaporation of water-rich ices (\cite{Loinard00} 2000; 
\cite{Ceccarelli01} 2001, \cite{Schoier02} 2002). Nevertheless, observation of abundant D$_2$CO 
in the outer envelop of IRAS16293 (\cite{Ceccarelli01} 2001) provides 
some observational support for onion-like mantle structure, with deuterated 
species trapped in the CO-rich ices that evaporate around 15K.  In this 
scheme gas-phase observation of deuterated molecules would trace the 
deuteration ratio in these CO-rich ices, whereas the solid-phase 
observation would trace the bulk of the H2O-ices, where deuteration may 
be less important.

Alternatively, water ices are formed by condensation
of water, copiously formed in molecular shocks, occuring during the
cloud phase (\cite{Bergin99} 1999). Also in this case the HDO/H$_2$O
ratio would be low, reflecting the high temperatures ($\sim 300$ K)
at which water would have been formed. Within this scenario, formaldehyde and
methanol are not shock-produced and hence show quite different (higher) 
deuterium fractionations characteristic of grain surface (or gas phase) 
formation at low temperatures during the preshock phase. Of 
course, shocks also have other chemical consequences - notably enhanced 
abundances of SiO and SO are considered to be common tracers of shocks 
(e.g., \cite{Codella02} 2002; \cite{Schilke97} 1997, and references therein). 
 However, this pertains to gas phase species and, moreover, cannot address the 
 deuteration issue at hand here.

None of these models seem to be fully developped and all seem to 
have some difficulty explaining the observations. Perhaps, 
 it is time to reconsider models for the selective deuteration of formaldehyde 
 and methanol on grain surfaces.

\section{Conclusions}

We presented a search for solid HDO in grain mantles towards low-mass
protostars. We did not detect HDO but derive upper limits on the HDO/H$_2$O 
frationation ratio of 0.5 to 2\%. These upper limits definitely show that 
solid water is much less deuterated than other molecules observed in the gas phase,
such as formaldehyde and 
methanol. The origin of these differences is not fully understood.

{} 

\end{document}